\documentclass[reprint,amsmath,amssymb,aps,prb,superscriptaddress,longbibliography]{revtex4-1}

\pdfoutput=1

\usepackage{hyperref,url}
\usepackage{amsmath}
\usepackage{amsfonts}
\usepackage[capitalise]{cleveref}
\usepackage{placeins}
\hypersetup{breaklinks,colorlinks=true,linkcolor=blue,citecolor=blue,urlcolor=blue,filecolor=blue}
\usepackage{graphicx}
\usepackage{dcolumn}
\usepackage{algorithm}
\usepackage{mhchem}
\usepackage{algpseudocode}
\usepackage{listings}
\usepackage{xcolor}

\definecolor{codegreen}{rgb}{0,0.6,0}
\definecolor{codegray}{rgb}{0.5,0.5,0.5}
\definecolor{codepurple}{rgb}{0.58,0,0.82}
\definecolor{backcolour}{rgb}{0.95,0.95,0.92}

\lstdefinestyle{mypython}{
    language=Python,
    backgroundcolor=\color{backcolour},   
    commentstyle=\color{codegray},
    keywordstyle=\color{magenta},
    numberstyle=\tiny\color{codegray},
    stringstyle=\color{codegreen},
    basicstyle=\ttfamily\footnotesize,
    breakatwhitespace=false,         
    breaklines=true,                 
    captionpos=b,                    
    keepspaces=true,                 
    numbers=left,                    
    morekeywords=[1]{einsum,dot,inv,empty,copy,reshape,zeros,shape,index,to_csv,run,kernel},
    numbersep=5pt,                  
    showspaces=false,                
    showstringspaces=false,
    showtabs=false,                  
    tabsize=2
}

\lstnewenvironment{python}[1][]{\lstset{style=mypython,#1}}{}

\Crefname{lstlisting}{Listing}{Listings}

\newcommand{\ipie}{\texttt{ipie}~}
\newcommand{\QMCPACK}{\texttt{QMCPACK}~}
\newcommand{\qmcpack}{\texttt{QMCPACK}}
\newcommand{\PAUXY}{\texttt{PAUXY}~}
\newcommand{\Dice}{\texttt{Dice}~}

\newcommand\mat\mathbf

\newcommand{\Google}{\affiliation{
Google Research, Venice, CA 90291, United States}}
\newcommand{\DeepMind}{\affiliation{DeepMind, 6 Pancras Square, London N1C 4AG}}
\newcommand{\Columbia}{\affiliation{Department of Chemistry, Columbia University, New York, NY, USA}}

\newcommand{\Boulder}{\affiliation{Department of Chemistry, University of Colorado, Boulder, CO 80302, USA}}

\begin{document}
\author{Fionn D. Malone}
\Google
\author{Ankit Mahajan}
\Boulder
\author{James S. Spencer}
\DeepMind
\author{Joonho Lee}
\email{linusjoonho@gmail.com}
\Columbia\Google

\title{\texttt{ipie}: A Python-based Auxiliary-Field Quantum Monte Carlo Program
with Flexibility and Efficiency on CPUs and GPUs
}

%\begin{document}
\begin{abstract}
We report 
the development of
a python-based auxiliary-field quantum Monte Carlo (AFQMC) program, \texttt{ipie}, with
preliminary timing benchmarks and new
AFQMC results on the isomerization of \ce{[Cu2O2]^{2+}}.
We demonstrate how implementations for both central and graphical processing units (CPUs and GPUs) are achieved in \texttt{ipie}.
We show an interface of \texttt{ipie} with \texttt{PySCF} as well as a straightforward template for adding new estimators to \texttt{ipie}.
Our timing benchmarks against other C++ codes, \texttt{QMCPACK} and \texttt{Dice}, suggest that \texttt{ipie} is faster
or similarly performing for all chemical systems considered on both CPUs and GPUs.
Our results on \ce{[Cu2O2]^{2+}} using selected configuration interaction trials 
show that it is possible to converge the ph-AFQMC isomerization energy between bis($\mu$-oxo) and $\mu$-$\eta^2$:$\eta^2$ peroxo configurations to the exact known results for small basis sets with $10^5$ to $10^6$ determinants. 
We also report the isomerization energy with a quadruple-zeta basis set with an estimated error less than a kcal/mol, which involved 52 electrons and 290 orbitals with $10^6$ determinants in the trial wavefunction.
These results highlight the utility of ph-AFQMC and \texttt{ipie} 
for systems with modest strong correlation and large-scale dynamic correlation.
\end{abstract}
\maketitle

\section{Introduction}
The development of
new electronic structure
methods
can often be 
accelerated by
programs written in
high-level, interpreted languages such as Python and Julia.
Prominent recent examples are \texttt{Psi4Numpy},\cite{Smith2018Jul}, \texttt{PySCF},\cite{Sun2020Jul} \texttt{PyQMC},\cite{pyqmc} \texttt{PyQuante},\cite{pyquante,pyquante2}  \texttt{JuliaChem.jl},\cite{Poole2020Aug,Poole2022Apr} \texttt{Fermi.jl},\cite{Aroeira2022Feb} and \texttt{iTensors.jl}.\cite{itensor}
If programs written in such high-level languages reach
similar performance
compared to other production-level programs written in Fortran or C++,
much more scientific development
can be achieved at a faster pace.
Motivated by this, in this work, 
we report our progress 
on developing a Python-based
auxiliary-field quantum Monte Carlo (AFQMC) 
code
that is both
performant
and flexible for new feature developments.

AFQMC 
has proven to be
one of the more promising
approaches to
many-electron correlation problems.\cite{Zhang2013,SimonsCollaborationontheMany-ElectronProblem2015Dec,Motta2018Sep,SimonsCollaborationontheMany-ElectronProblem2020Jul,Shi2021Jan,lee2022twenty}
Historically, 
the method has been mostly applied to lattice models within the constrained path approximation\cite{Zhang1995May}
but
there has been an increasing interest
in applying
this method within the phaseless approximation\cite{Zhang2003Apr}
to 
{\it ab initio} systems.\cite{Motta2018Sep,SimonsCollaborationontheMany-ElectronProblem2020Feb,Shi2021Jan,lee2022twenty}
Despite its success,
the open-source code development has seen
rather slow progress until recently although
a proof-of-concept implementation in \texttt{MATLAB}\cite{Nguyen2014Dec} was available
for the Hubbard model in 2014.
For {\it ab initio} systems,
it was only within the last 10 years when the first production-level, open-source implementation developed by Morales and co-workers in \QMCPACK appeared.\cite{kim2018qmcpack,Kent2020May}
Since then, another proof-of-concept program, \texttt{PAUXY},\cite{pauxy} written in Python by Malone and Lee, was made available.
Finally, Sharma and co-workers released their free-projection (fp-) and phaseless (ph-) AFQMC implementation as a part of their open-source software project, \texttt{Dice}.\cite{dice}

Our new code,\cite{ipie} distributed under the Apache 2.0 license, intelligent python-based imaginary-time evolution (\texttt{ipie}), was largely motivated by the
utility of \PAUXY recognized by two of us
after having used it for several research articles.\cite{Lee2019Aug,Lee2020Jul,Lee2021Feb,Lee2021Mar,Lee2021Jun}
Nonetheless, the full potential of \PAUXY was limited due to its computational inefficiency as well as its lack of graphics processing unit (GPU) support.
AFQMC has two computational hotspots: propagation and local energy evaluation.
The cost of these steps are dominated by dense linear algebra where the overhead of performing these in Python could be small.
Propagation is mainly matrix-matrix multiplication that can be performed on Python efficiently using \texttt{dot} operations in NumPy\cite{numpy} and CuPy.\cite{cupy}
Similarly, the local energy evaluation can be performed efficiently utilizing just-in-time (\texttt{jit}) compilation in Numba.\cite{numba}
Furthermore, other computationally light-weight parts such as walker orthogonalization and population control of AFQMC do not cause much overhead in Python.
These observations led us to develop \texttt{ipie}, an AFQMC code written in Python.
The key goal of this open-source project is to aim for both flexibility for development and efficiency for production-level calculations.

At the same time,
the recent development of quantum-classical hybrid ph-AFQMC algorithm\cite{Huggins2022Mar,Lee2022Jul}
started to
involve 
a new scientific community in AFQMC, namely the quantum information science (QIS) community. 
As many code developments in QIS have been dominated by Python,\cite{quantumlib2022Aug,Qiskit2022Aug,qsim,Rubin2021Oct,Stair2022Mar}
the development of a production-level python-based AFQMC code
has become even more relevant.
\ipie has been mostly written by
few core developers, but
we hope that our progress report here will attract
many scientists to this open-source software project.

Our paper is organized as follows.
First, we will briefly review the theory behind the core features of \texttt{ipie}.
Next, we will present some timing comparisons on both central processing units (CPUs) and GPUs against \QMCPACK and \Dice to showcase the performance of \texttt{ipie}.
As an example of developing new features,
we show how one can add a new mixed estimator 
to AFQMC without modifying internal codes in \texttt{ipie}.
Finally, we report
the AFQMC study of \ce{[Cu2O2]^{2+}}
with a heat-bath configuration interaction (HCI) trial wavefunction, compare with unbiased results,\cite{Mahajan2022May} and present highly accurate complete basis set limit results.

\section{Theory}
\subsection{Notation}
\begin{table}[h]
\begin{tabular}{c|c}
Notation                                & Description     \\\hline
MPI & Message Passing Interface \\\hline
($n$e,$m$o) & $n$ electrons in $m$ spatial orbitals \\\hline
$M$, nbasis & the number of orbitals\\\hline
$N$, nocc & the number of electrons\\\hline
$N_\text{aux}$, naux & the number of auxiliary vectors\\\hline
$N_\text{walkers}$, nwalkers & the number of walkers\\\hline
$N_\text{dets}$ & the number of determinants\\\hline
$p,q,r,s$, etc. & indices for orbitals \\\hline
$i,j,k,l$, etc. & indices for occupied orbitals \\\hline
$\alpha,\beta,\gamma$, etc. & indices for auxiliary vectors \\\hline
$a,b,c,d$, etc. &  indices for determinants\\\hline
$x,y,z$, etc. &  indices for walkers\\\hline
\end{tabular}
\caption{Notation used in this work.
\label{tab:notation}}
\end{table}
In \cref{tab:notation}, we summarize the notation used in the later sections.
We assume that we work in spin-orbital basis unless spins are labeled explicitly although our implementations write out spin degrees of freedom explicitly.

\subsection{Review of AFQMC}
In this paper, we hope to focus more on the computational and implementational aspects of ph-AFQMC as relevant to \texttt{ipie}.
We will only summarize the essence of AFQMC, so interested readers are referred to recent ph-AFQMC reviews\cite{Zhang2013,Motta2018Sep,Shi2021Jan,lee2022twenty} for more theoretical details on ph-AFQMC.

Projector QMC algorithms approximate the exact ground state, $|\Psi_{0}\rangle$, via imaginary time evolution:
\begin{equation}
|\Psi_{0}\rangle = \lim_{\tau \to \infty} e^{-\tau\hat{H}}|\Phi_{0}\rangle,
\label{eq:imag_proj}
\end{equation}
where $\tau$ is the imaginary time, $\hat{H}$ is the Hamiltonian,
and
$|\Phi_0\rangle$ is an initial wave function with non-zero overlap with $|\Psi_{0}\rangle$. 
Different QMC algorithms differ in how one implements the imaginary time propagation in \cref{eq:imag_proj}, how the Hamiltonian and initial wavefunction is represented and how the method controls the fermionic sign (or phase) problem.

In \texttt{ipie}, we implemented a specific flavor of QMC, namely, AFQMC.
In AFQMC, we work with a second quantized representation of the Hamiltonian
\begin{equation}
\hat{H}
=
\sum_{pq}h_{pq}a_p^\dagger a_q
+ \frac12 \sum_{pqrs}
(pr|qs) a_p^\dagger a_q^\dagger a_s a_r,
\label{eq:Habinitio}
\end{equation}
and assume that the two-electron integral tensor is factorized by the Cholesky or density-fitting factorization
\begin{equation}
(pr|qs) = \sum_{\alpha=1}^{N_\text{aux}} L_{pr}^\alpha L_{qs}^\alpha,
\end{equation}
with $L_{pr}^\alpha$ being auxiliary vectors.
In AFQMC, one further discretizes $\tau$ and applies the Trotter decomposition \cite{Trotter1959Aug} followed by the Hubbard-Stratonovich (HS) \cite{hubbard_strat} transformation to the propagator $e^{-\Delta \tau\hat{H}}$. 
This then leads to the integral representation of the imaginary-time propagator for a small time step, $\Delta \tau$,
\begin{equation}
e^{-\Delta \tau\hat{H}} = \int \mathrm{d}^{N_\text{aux}}\textbf{x}  p(\textbf{x}) \hat{B}(\textbf{x}) + \mathcal O (\Delta \tau^2),
\label{eq:Bx}
\end{equation}
where $p(\textbf{x})$ is the normal distribution and $\hat{B}(\textbf{x})$ is
a one-body propagator coupled to
a vector of $N_\text{aux}$ auxiliary fields
\textbf{x}. 

From here the AFQMC algorithm could proceed by evolving a set of walkers by sampling the application of the propagator in \cref{eq:Bx} and updating the walker weights accordingly.
This free-projection algorithm (fp-AFQMC) is however plagued by the fermion phase problem\citep{Zhang2013}.
Although using an accurate initial state and trial wavefunction can delay the onset of the phase problem, ultimately it limits the applicability of fp-AFQMC to relatively short projection times (typically 1-2 a.u.), a situation which only gets worse with system size.

To overcome this, Zhang et al. introduced the phaseless approximation
 \cite{Zhang2003Apr} (ph-AFQMC hereafter)  utilizing importance sampling based on an a priori chosen trial wavefunction, $|\Psi_T\rangle$ in addition to a fixing of the walker's phase which is approximately enforced through the trial wavefunction.
 With importance sampling, the global wave function at time $\tau$ is written as a weighted statistical sum over $N_\text{walkers}$ walkers,
\begin{equation}
|\Psi(\tau)\rangle = 
\sum_{z=1}^{N_\text{walkers}}
w_z (\tau)
\frac{|\phi_z(\tau)\rangle}
{\langle \Psi_T|\phi_z(\tau)\rangle}.
\label{eq:stat_wfs}
\end{equation}
where $\{\phi_z(\tau)\}$ consists of single Slater determinants in arbitrary single-particle basis.
ph-AFQMC ensures the positivity of walker weights $\{w_z (\tau)\}$ at all time $\tau$ and is thereby statistically efficient (i.e., the variance only grows linearly with system size), albeit at the expense of introducing a systematically improvable bias.
We note that systematically improving the bias is generally thought to be exponentially expensive, similar to other constrained projector Monte Carlo approaches. 

This importance sampling transformation yields the `hybrid method' for propagating walkers:
\begin{align}\nonumber
&w_z(\tau+\Delta\tau) |\phi_z(\tau+\Delta\tau)\rangle =\\
&\left[
I(\mathbf{x}_z,\mathbf{\bar{x}}_z,\tau,\Delta\tau)
\hat{B}(\Delta \tau, \mathbf {x}_z-\mathbf{\bar{x}}_z)
\right] w_z(\tau)|\phi_z(\tau)\rangle,
\end{align}
where the importance function is defined as 
\begin{equation}
    I(\mathbf{x}_z,\mathbf{\bar{x}}_z,\tau,\Delta\tau) = O_z(\tau, \Delta\tau)
    e^{\mathbf{x}_z\cdot\mathbf{\bar{x}}_z-\mathbf{\bar{x}}_z\cdot\mathbf{\bar{x}}_z/2},
 \label{eq:import}
\end{equation}
$O_z$ is the overlap ratio
\begin{equation}
O_z(\tau, \Delta\tau) = \frac{\langle
    \Psi_T |
    \hat{B}(\Delta\tau, \mathbf{x}_z-\mathbf{\bar{x}}_z) | \phi_z(\tau)
    \rangle}{
    \langle
    \Psi_T | \phi_z(\tau)
    \rangle},
 \label{eq:ovl}
\end{equation}
and $\mathbf{\bar{x}}_z$  is the optimal force bias, given as
\begin{equation}
    \mathbf{\bar{x}}_z(\Delta \tau, \tau) = -\sqrt{\Delta\tau}
\frac{\langle
    \Psi_T | \hat{\mathbf{v}}' | \phi_z(\tau)
    \rangle}{
    \langle
    \Psi_T | \phi_z(\tau)
    \rangle
    }.
\end{equation}
The phaseless approximation (ph) is then defined as a modification to this importance function
\begin{equation}
I_{ph}(\mathbf{x}_z,\mathbf{\bar{x}}_z,\tau,\Delta\tau) = |I (\mathbf{x}_z,\mathbf{\bar{x}}_z,\tau,\Delta\tau)|\times
\text{max}(0, \cos(\theta_z(\tau)))
\label{eq:ph}
\end{equation}
where the phase $\theta_z(\tau)$ is given by
\begin{equation}
\theta_z(\tau) = \text{arg}\left(
O_z(\tau, \Delta\tau)
\right).
\label{eq:theta}
\end{equation}
The walker weights and Slater determinants are then updated as 
\begin{align}
w_z(\tau+\Delta\tau) &= I_{ph}(\mathbf{x}_z,\mathbf{\bar{x}}_z,\tau,\Delta\tau) \times w_z(\tau)\\
|\phi_z(\tau+\Delta\tau)\rangle &= \hat{B}(\Delta\tau, \mathbf{x}_z-\mathbf{\bar{x}}_z) |\phi_z(\tau)\rangle. 
\end{align}

\section{Implementational details}
In this section, we focus on three computational kernels that are at the core of ph-AFQMC and discuss how they are implemented.
The general strategy that we take is to
avoid looping over
walkers as much as we can
because loops often
involve additional overheads in Python.

We currently support two types of trial wavefunctions, $|\Psi_T\rangle$: single- and multi-Slater determinant (SD and MSD) wavefunctions,
\begin{equation}
|\Psi_T\rangle
=
\sum_{a=0}^{N_\text{det}-1} c_a |\psi_a\rangle
=
\sum_{a=0}^{N_\text{det}-1} c_a \hat{t}_a |\psi_0\rangle
\end{equation}
where $c_a$ is the expansion coefficient (sorted in descending order) for the $a$-th determinant, and
$\hat{t}_a$ is the $a$-th excitation operator that generates $|\psi_a\rangle$ from $|\psi_0\rangle$ (a reference determinant).
While our implementation for SD is relatively straightforward, the implementation for $N_\text{det}>1$ (i.e., MSD) is more involved
following the
generalized Wick's theorem.\cite{Mahajan2021Aug,Mahajan2022May}
We assume that the determinants, $|\psi_a\rangle$, are mutually orthonormal and the corresponding coefficient matrices for each determinant are represented over orthogonal basis sets.
\ipie currently supports SD trials for both CPU and GPU and MSD trials for only CPU.

\subsection{Overlap}
In ph-AFQMC, one needs to compute the overlap between trial and walker wavefunctions:
\begin{equation}
\langle\Psi_T |\phi_z \rangle
= 
\sum_{a=0}^{N_\text{det}-1}
c_a\langle\psi_a|\phi_z\rangle
\end{equation}
For SD trials ($N_\text{det}=1$), this involves straightforward linear algebra function calls:
\begin{equation}
\langle\psi_0 |\phi_z \rangle    
=
\det
\left(
\mathbf C^\dagger_{\psi_0} \mathbf C_{\phi_z}
\right)
=
\det
\left(
\mathbf C_{\phi_z}^T
\mathbf C^*_{\psi_0} 
\right)
\end{equation}
where
$\mathbf C_{\psi_0}$ and $\mathbf C_{\phi_i}$ are coefficient matrices for
$|\psi_0\rangle$ and $|\phi_i\rangle$, respectively.
We implement this operation for both CPU and GPU by using an \texttt{einsum} call followed by a determinant evaluation:
\begin{python}[caption=Code snippet for constructing overlap on CPU and GPU.]
S = einsum("zpi,pj->zij", phi, psi.conj())
overlap = linalg.det(S)
\end{python}
where \texttt{einsum} and \texttt{linalg.det} are using NumPy and CuPy implementations for CPU and GPU, respectively.
We note that walker coefficient matrices are stored as a three-dimensional array with size of $N_\text{walkers}\times M \times N$. 
This then allows for a single \texttt{einsum} call for computing the dot product between two wavefunctions.
The cost of this operation scales as $\mathcal O(N_\text{walkers} M N^2)$ in general cases although the cost reduction to $\mathcal O(N_\text{walkers} M N)$ is possible in the molecular orbital basis (which is typical for MSD trials).
The subsequent determinant evaluation costs $\mathcal O(N_\text{walkers} N^3)$.

For MSD trials, the code is much more involved. 
We compute the overlap by
\begin{equation}
\langle \Psi_T | \phi_z\rangle =
    \langle \psi_0 | \phi_z\rangle + \langle \psi_0 | \phi_z\rangle \left(\sum_{a=1}^{N_\text{det}-1} c_a \frac{\langle \psi_0 |\hat{t}_a^\dagger| \phi_z\rangle}{\langle \psi_0 | \phi_z\rangle}\right).
\label{eq:msdovlp}
\end{equation}
The summation over determinants in the second term is performed via the generalized Wick's theorem\cite{Mahajan2021Aug} that uses the following quantity as an intermediate:
\begin{equation}
G_{pq}^{0,z} = \frac{\langle \psi_0 | 
\hat{a}_p^\dagger \hat{a}_q
|\phi_z\rangle}{\langle \psi_0 | \phi_z\rangle}
\label{eq:msdgf0}
\end{equation}
where the upper index 0 refers to the first determinant index.
The implementation to compute this intermediate will be discussed below.
%Overall, the overlap evaluation for MSD trials costs
%$\mathcal O(N_\text{walkers}N_\text{det}k^3 + N_\text{walkers}MN + N_\text{walkers}N^3)$ where $k$ is the typical excitation rank in $|\Psi_T\rangle$.
Implementation details on this algorithm for GPU will be reported in the future work.

\subsection{One-body Green's function and force bias}
The most common propagation algorithm for ph-AFQMC is called the hybrid algorithm.\cite{Zhang2003Apr,Purwanto2004Nov}
A key component this algorithm involves computing the so-called optimal ``force bias'' potential defined as
\begin{align}\nonumber
\bar{x}_\alpha^z &= \frac{\langle \Psi_T | a_p^\dagger a_q | \phi_z\rangle}{\langle\Psi_T|\phi_z\rangle}L_{pq}^\alpha\\
&=\sum_{pq}G_{pq}^z L_{pq}^\alpha, \label{eq:fb}
\end{align}
where
the one-body Green's function reads
\begin{equation}
G_{pq}^z = \frac{\langle \Psi_T | 
\hat{a}_p^\dagger \hat{a}_q
|\phi_z\rangle}{\langle \Psi_T | \phi_z\rangle}.
\label{eq:gf}
\end{equation}
The force bias evaluation is a simple dot product so \cref{eq:fb} can be naively performed at the cost of $\mathcal O(N_\text{walkers}N_\text{aux}M^2)$.

It is helpful for computational efficiency when constructing the force bias and also evaluating the local energy
to construct the following intermediate 
tensor by transforming one-index of auxiliary vectors:
\begin{equation}
L_{ir}^\alpha
=
\sum_{p} (\mathbf C_{\psi_0}^\dagger)_{ip} L_{pr}^\alpha.
\end{equation}
This is stored in shared memory as
a $N_\text{aux}\times NM$ matrix in our code and only needs to be constructed once in the beginning of the simulation (per computational node).

For SD trials this allows us to exploit a  
rank-revealing, factorized  form for the Green's function,
\begin{equation}
G_{pq}^z = 
(\mathbf C_{\psi_z} (\mathbf C_{\psi_0}^\dagger \mathbf C_{\psi_z})^{-1} \mathbf C_{\psi_0}^\dagger)_{qp}
=
\sum_{j=1}^{N}
\theta^z_{jp}
(\mathbf C_{\psi_0}^\dagger)_{jq},
\label{eq:halfgf}
\end{equation}
where the walker-dependent part of the Green's function is
\begin{equation}
\theta^z_{jp}
=
(\mathbf C_{\psi_z} (\mathbf C_{\psi_0}^\dagger \mathbf C_{\psi_z})^{-1})_{pj},
\end{equation}
so, for example, we can write \cref{eq:fb} as 
\begin{equation}
\bar{x}_{\alpha}^z = 
\sum_{pj}\theta_{jp}^z L_{pj}^\alpha.
\label{eq:rfb}
\end{equation}
The cost of building the force bias in this form is $\mathcal O(N_\text{walkers}N_\text{aux}MN)$.
In our GPU implementation,
we build $\theta_{jp}^z$ by
\begin{python}[caption=Code snippet for constructing $\theta_{jp}^z$ on GPU.]
import cupy as cp # For GPU implementation
S = cp.einsum("zpi,pj->zij", phi, psi.conj())
Sinv = cp.inv(S)
theta = cp.einsum("zji,zpi->zjp", Sinv, phi)
\end{python}
For CPU, we found that it is faster to loop over the walker index and perform \texttt{dot} and \texttt{inv} operations than using \texttt{einsum} and batched \texttt{inv}.
Using $\theta_{jp}^z$, 
the force bias can be built as a matrix-matrix multiplication for all walkers in a given MPI task at once:
\begin{python}[caption={Code snippet for constructing force bias on CPU and GPU.},label={list:fb}]
theta = theta.reshape((nwalkers, nocc*nbasis))
force_bias_real = rchol.dot(theta.T.real)
force_bias_imag = rchol.dot(theta.T.imag)
force_bias = empty((nwalkers, naux),
                    dtype=theta.dtype)
force_bias.real = force_bias_real.T.copy()
force_bias.imag = force_bias_imag.T.copy()
\end{python}
where \texttt{rchol} denotes $L_{pj}^\alpha$ with a shape of ($N_\text{aux}$, $NM$).
We use the same code for both CPU and GPU where \texttt{empty} is from NumPy for CPU and from CuPy for GPU. 
We found it more efficient to compute real and imaginary parts
of the force bias separately. Na{\"i}ve \texttt{dot} operations between a real-valued array \texttt{rchol} and a complex-valued array \texttt{theta} create a complex-valued copy for \texttt{rchol}, which consumes additional memory and slows down the overall operation. Therefore, we avoid this by creating copies.

For MSD trials, we compute the denominator of \cref{eq:gf} using \cref{eq:msdovlp} and the numerator via the generalized Wick's theorem:\cite{Mahajan2021Aug}
\begin{equation}
\langle \Psi_T |\hat{a}_p^\dagger \hat{a}_q|\phi_z\rangle
=
\langle \psi_0 | \phi_z\rangle \left(\sum_{a=0}^{N_\text{det}-1} c_a \frac{\langle \psi_0 |\hat{t}_a^\dagger \hat{a}^\dagger_p \hat{a}_q |\phi_z\rangle}{\langle \psi_0 | \phi_z\rangle}\right).
\end{equation}
The necessary intermediate in \cref{eq:msdgf0} can be computed by \cref{eq:halfgf}.
We largely followed the CPU implementation described in Ref. \citenum{Mahajan2022May} with some necessary changes to avoid excessive looping due to Python overheads. 
We will report more detailed descriptions about it in our forthcoming GPU implementation paper.

\subsection{Local energy}
The local energy is one of the most important mixed estimators
to compute in QMC methods.
It is defined as
\begin{equation}
E^z_L = 
\frac
{\langle \Psi_T | \hat{H} | \phi_z\rangle}
{\langle \Psi_T | \phi_z\rangle},
\end{equation}
and the global energy (i.e. walker averaged) estimate can then be computed by
\begin{equation}
E = \frac{\sum_z w_z E_L^z}{\sum_z w_z}
\end{equation}
There are many algorithms developed for the local energy evaluation\cite{Motta2019Jun,Malone2019Jan,Lee2020Jul,Weber2022Jun} and we implemented some of the more commonly used algorithms.

For brevity, we focus on the exact exchange-like energy contribution as it is more computationally demanding (i.e., quartic-scaling) than that of Coulomb-like one (i.e., cubic-scaling.)
For SD trials, this contribution reads
\begin{equation}
E^z_{L,\text{K}}
=
-\frac12\sum_{ij}\sum_{pq}\sum_\alpha L_{ip}^\alpha L_{jq}^\alpha \theta_{iq}\theta_{jp}.
\end{equation}
Looping over $\alpha$ and contracting over other indices to accumulate the energy contribution will scale as
$\mathcal O(N_\text{walkers}N_\text{aux}N^2M)$. The computational bottleneck in this step is to form
\begin{equation}
T_{ij}^{z,\alpha} = \sum_p L_{ip}^\alpha \theta_{jp}^z.
\end{equation}
To avoid storing this object in memory, we use a direct algorithm where
$T_{ij}^{z,\alpha}$ for each of $z$ and $\alpha$ is formed and consumed immediately.
Looping over both walker and auxiliary indices creates additional overheads in Python so we resort to Numba and its \texttt{jit} compilation to remove such overheads:
\begin{python}[caption={Code snippet for computing exchange-like contribution to the local energy. It is written for Numba's \texttt{jit} capability.}]
import numba.jit as jit
import numpy as np
@jit(nopython=True, fastmath=True)
def compute_exchange(rchol, theta):
    naux = rchol.shape[0]
    nwalkers = theta.shape[0]
    nocc = theta.shape[1]
    nbasis = theta.shape[2]
    T = np.zeros((nocc, nocc),dtype=np.complex128)
    exx = np.zeros((nwalkers),dtype=np.complex128)
    rchol = rchol.reshape((naux, nocc, nbasis))
    for iw in range(nwalkers):
        theta_r = theta[iw].T.real.copy()
        theta_i = theta[iw].T.imag.copy()
        for jx in range(naux):
            T = (
                rchol[jx].dot(theta_r)
                +
                1.0j * rchol[jx].dot(theta_i)
              )
            exx[iw] += np.dot(T.ravel(),
                       T.T.ravel())
    exx *= 0.5
    return exx
\end{python}
In the above code, we split real and imaginary parts of $\theta_{jp}^z$ by creating appropriate copies (line 13-14). This was found to be helpful in performing subsequent \texttt{dot} operations as these copies store real and imaginary elements of $\theta_{jp}^z$ contiguously in memory. Furthermore, it avoids creating a complex-valued copy of \texttt{rchol} as explained before for \cref{list:fb}.
For our GPU implementation, we found that utilizing only existing CuPy functions
leads to
either inefficient or very memory-demanding local energy evaluations.
It was therefore necessary to write our own GPU kernel with Numba's \texttt{cuda.jit} based on the same local energy algorithm implemented in \texttt{QMCPACK}.\citep{Kent2020May}

For MSD trials, we utilize the generalized Wick's theorem similarly to how overlap and Green's functions were evaluated. For one-body energies, we can easily compute the contribution by using the one-body Green's function in \cref{eq:gf}.
Additional complications arise for the two-body local energy contribution where
we evaluate the numerator as
\begin{align}\nonumber
\langle\Psi_T|\phi_z\rangle E_{L,2}^z = &-\frac12
\langle \psi_0 | \phi_z\rangle 
\sum_{prqs}
\sum_{\alpha=1}^{N_\text{aux}}
\sum_{a=0}^{N_\text{det}-1} \\
&\bigg(
c_a \frac{\langle \psi_0 |\hat{t}_a^\dagger\hat{a}^\dagger_p \hat{a}_r \hat{a}_q^\dagger \hat{a}_s |\phi_z\rangle}{\langle \psi_0 | \phi_z\rangle}
L_{pr}^\alpha
L_{qs}^\alpha
\bigg).
\end{align}
We extensively used Numba's \texttt{jit} capability and exploit \texttt{BLAS} for implementing the necessary loops and contractions for the generalized Wick's theorem over determinants, walkers, and excitations.
More detailed description about this implementation and its GPU extension will be reported in our forthcoming paper.

\subsection{Code Structure and Development Process}
\begin{figure*}[!ht]
    \centering
    \includegraphics[scale=0.5]{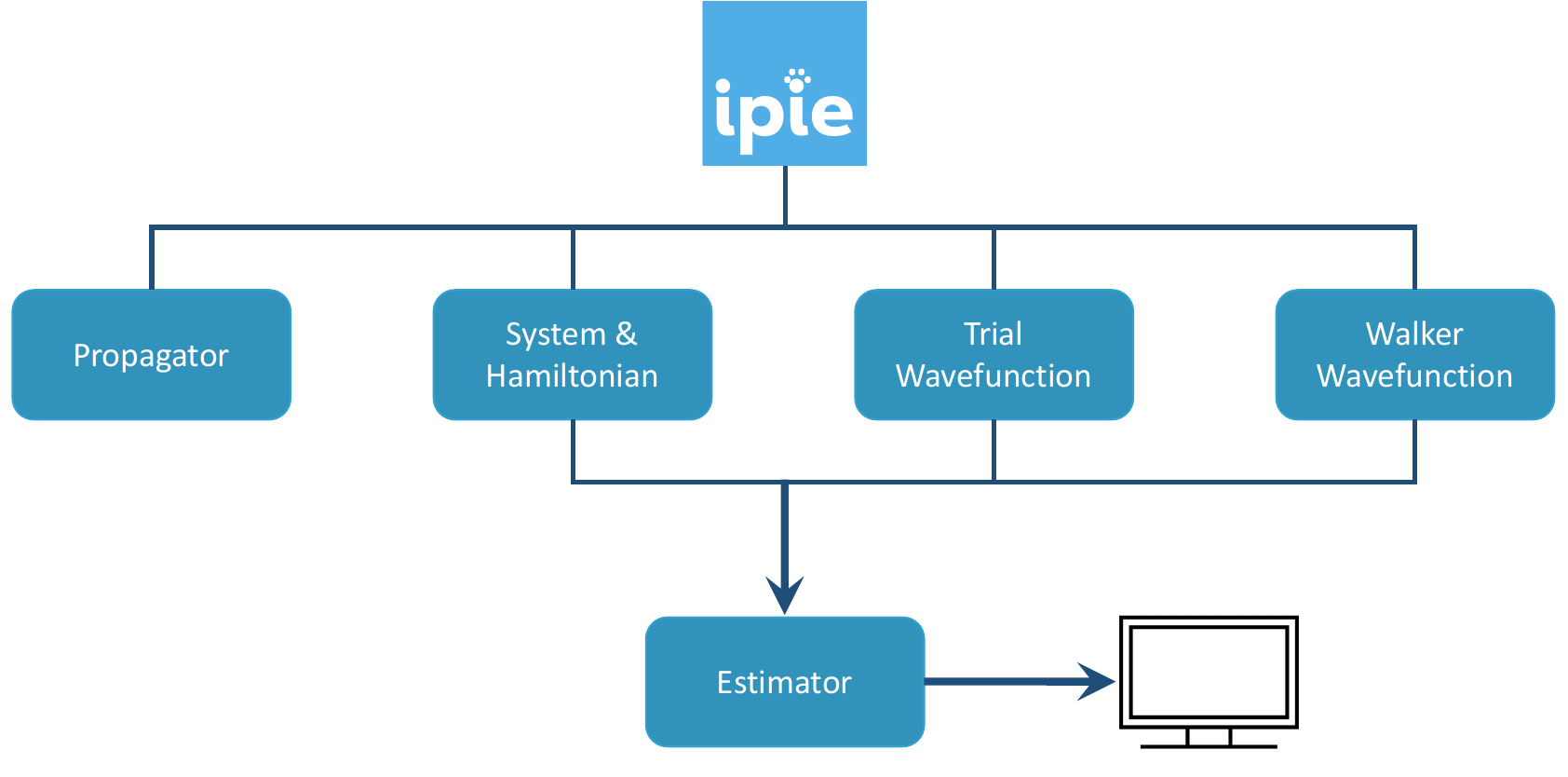}
    \caption{Diagram on how different classes are organized in \texttt{ipie}.}
    \label{fig:diagram}
\end{figure*}
In \cref{fig:diagram}, we present an overview of the internal structure of \texttt{ipie}.
We have seven distinct classes in \texttt{ipie} for {propagator}, {system}, {Hamiltonian}, {trial} wavefunction, walker wavefunction, and estimator.
The propagator class interacts with all other classes except estimator to evolve the walker wavefunction.
The system class specifies the number of electrons, symmetry of the system, and other information pertinent to the problem. 
The Hamiltonian class stores necessary one-body matrix elements and auxiliary vectors. 
The trial wavefunction class contains information about $|\Psi_T\rangle$ along with necessary integral intermediates that depend on $|\Psi_T\rangle$.
The walker wavefunction class stores weights and $|\phi_z\rangle$.
The estimator class takes in system, Hamiltonian, trial wavefunction, and walker wavefunction to produce the global observables at a given imaginary time.
We designed the program in a modular way to maximize the flexibility of adding new features in the future.

When adding new features to the code we try to follow a test driven development process, with new features initially prototyped as unit tests before being fully incorporated into the code. In addition to unit tests we have a set of deterministic serial and parallel integration tests designed to test the full AFQMC application.
The test suite is run through continuous integration upon a pull request being opened on GitHub with one approving code review required before any new feature or fix is merged into the development branch.
\section{Applications and Discussion}
In this section,
we hope to demonstrate
the ease of running our program when interfaced with \texttt{PySCF}\cite{Sun2020Jul}
and adding 
a new estimator $\langle\hat{S}^2\rangle$
as well as
a new application of ph-AFQMC to 
the computation of isomerization energy of \ce{[Cu2O2]^{2+}}.

\subsection{Interface with \texttt{PySCF}}
\ipie is a Python program, and thus the
most natural quantum chemistry package to be
paired with for obtaining integrals 
is currently \texttt{PySCF}.
While our interface is generic enough to be coupled with any quantum chemistry software such as \texttt{Q-Chem}\cite{Epifanovsky2021Aug} as long as users provide integrals in an expected format,
we expect that our biggest user base at the point of writing will be familiar with \texttt{PySCF}.

An AFQMC calculation on one-dimensional \ce{H10} (bond distance of 1.6 Bohr) in the STO-6G basis with a Hartree-Fock trial wavefunction can be performed by running a script like the following:
\begin{python}[caption={Script that runs a Hartree-Fock calculation using \texttt{PySCF}, generates relevant \texttt{ipie} inputs, runs a ph-AFQMC calculation via \texttt{ipie}, and analyzing the output of \texttt{ipie} for the final ph-AFQMC energy and its error bar.}]
from pyscf import gto, scf
from mpi4py import MPI
from ipie.utils.from_pyscf import (
            gen_ipie_input_from_pyscf_chk
            )

comm = MPI.COMM_WORLD
mol = gto.M(
        atom=[("H", 1.6 * i, 0, 0)\
              for i in range(0, 10)],
        basis="sto-6g",
        verbose=4,
        unit="Bohr",
    )
nelec = mol.nelec
if comm.rank == 0:
    mf = scf.UHF(mol)
    mf.chkfile = "scf.chk"
    mf.kernel()
    gen_ipie_input_from_pyscf_chk(mf.chkfile,
                                  verbose=0)
comm.barrier()

from ipie.qmc.calc import build_afqmc_driver
afqmc = build_afqmc_driver(comm, nelec=nelec, 
                      nwalkers_per_task=1000)

afqmc.qmc.nblocks = 200
afqmc.run(comm=comm)

if comm.rank == 0:
    from ipie.analysis.extraction import (
        extract_observable
        )
    fname = afqmc.estimators.filename
    qmc_data = extract_observable(fname,
                                "energy")
    y = qmc_data["ETotal"]
    y = y[50:] # discard first 50 blocks

    from ipie.analysis.autocorr import ( 
        reblock_by_autocorr
        )
    df = reblock_by_autocorr(y)
    print(df.to_csv(index=False))
\end{python}
This script uses 1000 walkers per MPI process to sample 200 AFQMC blocks, where a block consists of a number of steps where the propagator is applied (typically 25) and one energy evaluation.
Projector Monte Carlo methods generally produce serially correlated data and thus it is necessary to account for the autocorrelation time through approximately estimating it or using a reblocking analysis.\citep{flyvbjerg1989error}
Thus, we next discard 50 blocks for equilibration and reblock the remaining 150 blocks to produce an estimate of the energy and its error bar. The resulting energy is -5.3825(7) which is comparable to the energy of CCSD(T), -5.3841.

Note, the above example is for demonstration purposes only. For large scale production calculations it is not advised to run the initial SCF step in the same script as the subsequent AFQMC calculation, as often one needs to use hundreds or thousands of MPI processes and thus it is not a good use of computational resources, and we advise users to run the QMC workflow in distinct steps. 

\subsection{Customized mixed estimators}
Often, the total energy is not all that one wants to compute from ph-AFQMC simulations.
Ideally, adding arbitrary mixed estimators or even back-propagated estimators\cite{Motta2017Nov} should be
possible without much effort.
As an example of how this is accomplished in \texttt{ipie}, 
we show an implementation of $\langle \hat{S}^2\rangle$.
Since $[\hat{S}^2, \hat{H}] = 0$,
a mixed estimator for $\hat{S}^2$ is unbiased.
For an SD trial ($|\Psi_T\rangle =|\psi_0\rangle$), it can be computed by
\begin{align}\nonumber
\frac{\langle\psi_0 |\hat{S}^2|\phi_z\rangle}
{\langle\psi_0|\phi_z\rangle}
&=
N_\downarrow + M_S(M_S+1)\\
&-\sum_{pq}
G^z_{p_\downarrow q_\downarrow}
G^z_{p_\uparrow q_\uparrow}
\end{align}
where
$N_\downarrow$ is the number of spin-$\downarrow$ electrons and
$M_S$ is the $\hat{S}_z$ quantum number of the state.
$G^z_{p_\downarrow q_\downarrow}$ and
$G^z_{p_\uparrow q_\uparrow}$ are
one-body Green's functions (\cref{eq:halfgf}) for each spin section.

To add this new estimator to \texttt{ipie}, one can define a pertinent estimator class and add it to the driver class to run ph-AFQMC as the following:
\begin{python}[caption={Code snippet for implementing a new mixed estimator, $\langle \hat{S}^2\rangle$ and running \texttt{ipie} with it.}]
import numpy as np
from ipie.estimators.estimator_base import (
    EstimatorBase
    )
from ipie.estimators.greens_function_batch import (
    greens_function
    )

class S2Mixed(EstimatorBase):
    def __init__(self, ham):
        self._data = {
            "S2Numer": np.zeros((1), 
            dtype=np.complex128),
            "S2Denom": np.zeros((1), 
            dtype=np.complex128),
        }
        # shape of the estimator
        self._shape = (1,)

        # optional variables
        # whether to print into text output
        self.print_to_stdout = True
        # new filename if needed
        self.ascii_filename = None
        # whether to store it as scalar
        self.scalar_estimator = False

    def compute_estimator(self, sys, walkers,
                          hamiltonian, trial):
        greens_function(walkers, trial,
                        build_full=True)

        ndown = sys.ndown 
        nup = sys.nup
        Ms = (nup-ndown)/2.0

        two_body = -np.einsum("zij,zji->z",                       
                   walkers.Ga,walkers.Gb)
        two_body = two_body * walkers.weight

        denom = np.sum(walkers.weight)
        numer = np.sum(two_body)\ 
              + denom*(Ms*(Ms+1)+ndown)

        self["S2Numer"] = numer
        self["S2Denom"] = denom


# ... Build afqmc inputs and driver class  
# afqmc is the driver class
ham = afqmc.hamiltonian
afqmc.estimators["S2"] = S2Mixed(ham=ham)
afqmc.run(comm=comm)
\end{python}
The design uses inheritance with the custom estimator class deriving from the base \texttt{EstimatorBase} class which defines generic functionality common to all estimators. In this way complications such as MPI are abstracted away from the developer and only core functionality needs to be implemented.
As this code demonstrates,
adding new mixed estimators to
\texttt{ipie} 
does not require
modifying any code inside \texttt{ipie}
lowering the
barrier for prototyping
research related to estimators.
In the future, we hope to support
other ways to compute estimators such as
backpropagation\cite{Motta2017Nov}
or automatic differentiation.\cite{Sorella2010Dec}

\subsection{Timing benchmarks}
One of the common concerns
about writing a Python code for
many-body methods
is that it may be difficult for a Python-based code to
perform as well as those written in C++ or Fortran.
Therefore, in this section, we report some preliminary timing benchmarks comparing \texttt{ipie} against two production-level C++ codes, \texttt{QMCPACK}\cite{Kent2020May} and \texttt{Dice}.\cite{dice}

\begin{figure}[!h]
    \centering
    \includegraphics{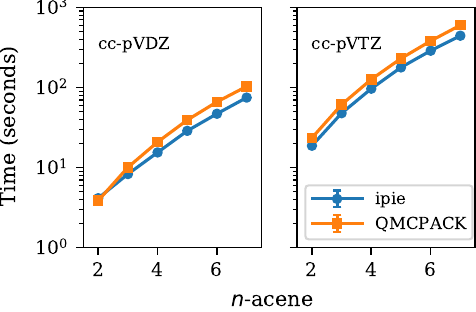}
    \caption{
    Average block time comparison between \ipie and \QMCPACK for the acenes series in the cc-pVDZ (left panel) and cc-pVTZ (right panel) basis sets. Calculations were performed on a single CPU (Intel Xeon @ 3.1 GHZ) and used 10 walkers. Timings were averaged across five independent runs each of which performed ten blocks. A block consisted of 25 propagation steps and one energy evaluation and orthogonalization. Standard error bars are plotted but are smaller than the points.
    }
    \label{fig:acenes_cpu_block}
\end{figure}
\begin{figure}[!h]
    \centering
    \includegraphics{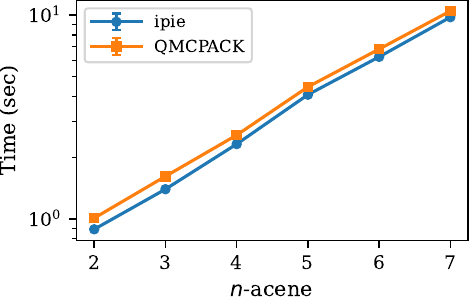}
    \caption{Average GPU block time comparison between \ipie and \QMCPACK for acenes series in the cc-pVDZ basis set. Calculations were performed on a single V100 GPU with 100 walkers. A block consisted of 25 propagation steps and one energy evaluation and orthogonalization. Timings were averaged five independent runs each of which performed ten blocks. Standard error bars are plotted but are smaller than the points.}
    \label{fig:acenes_gpu_block}
\end{figure}
\begin{figure}[!h]
    \centering
    \includegraphics{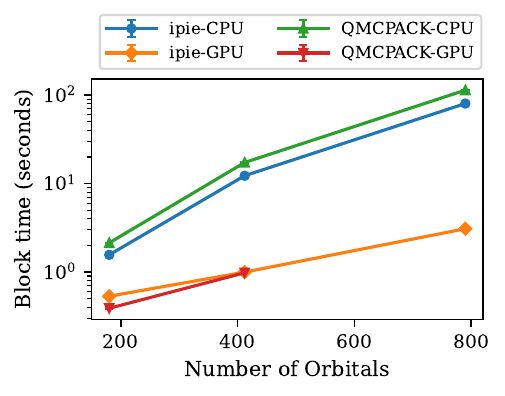}
    \caption{Average GPU block time comparison between \ipie and \QMCPACK for naphthalene as a function of basis set size (cc-pVDZ, cc-pVTZ, cc-pVQZ). GPU calculations were performed on a single V100 GPU with 30 walkers and CPU calculations were run with 30 MPI-processes (Intel Xeon @ 3.1 GHZ). A block consisted of 25 propagation steps and one energy evaluation and orthogonalization. Timings were averaged five independent runs each of which performed ten blocks. Standard error bars are plotted but are smaller than the points. QMCPACK would not run on a single V100 without distribution of integrals (a V100 has a memory limit of 16 GB) and thus we could not run it for comparison. ipie exploits symmetry in the Cholesky tensor to reduce its memory footprint, which delays the need for redistribution.}
    \label{fig:naphthalene_basis_set}
\end{figure}

In \cref{fig:acenes_cpu_block},
we compare the CPU performance of
\texttt{ipie} with SD trials for studying 
acene series from naphthalene to heptacene with cc-pVDZ and cc-pVTZ bases against \qmcpack.
For both bases, we see that \ipie outperforms \texttt{QMCPACK} for all system sizes except naphthalene in cc-pVDZ.
Similarly in \cref{fig:acenes_gpu_block},
we compare the GPU performance of
\texttt{ipie} with SD trials for studying 
acene series with cc-pVDZ basis against \qmcpack.
\ipie is faster than \texttt{QMCPACK} for all acenes.
Finally, in \cref{fig:naphthalene_basis_set} we compare timings as a function of basis sets size on a single CPU node (30 MPI processes) and a single V100, both with 30 walkers total.
Again we find ipie performs favorably relative to QMCPACK and we see a roughly order of magnitude gain in speed on the GPUs. 
We find that the ipie is typically slower than QMCPACK on GPUs when an insufficient number of walkers are used for small system sizes (indicating the the GPU is not being saturated), which is clear when comparing the cc-pVDZ naphthalene times in \cref{fig:acenes_gpu_block} and \cref{fig:naphthalene_basis_set}.
These timing benchmarks emphasize the utility of \texttt{ipie} beyond prototyping and
shows its potential for
production-level AFQMC calculations.

\begin{figure}[!h]
    \centering
    \includegraphics{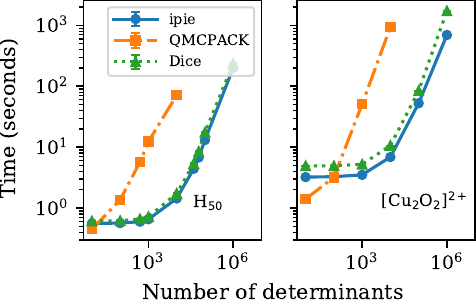}
    \caption{Comparison of the average time to compute one AFQMC block consisting of 25 propgation steps and one energy evaluation step on a single CPU core with \qmcpack, \texttt{ipie}, and \Dice codes for \ce{H50} for STO-6G and \ce{[Cu2O2]^{2+}} for the BS1 basis\cite{Mahajan2022May} as a function of the number of determinants in the trial wavefunction. Note that \texttt{QMCPACK} uses a different algorithm (i.e., Sherman-Morrison) than \ipie and \Dice. Results were averaged across 5 independent simulations (with a fixed seed) with each simulations performing ten blocks.}
    \label{fig:cu2o2_timing}
\end{figure}

We also benchmarked our CPU implementation of the generalized Wick's theorem as shown in \cref{fig:cu2o2_timing}.
\texttt{QMCPACK} employs a different algorithm based on the Sherman-Morrison formula, which inevitably becomes too slow to run for large determinant counts.\cite{Shee2018Aug,Mahajan2022May}
However, both \texttt{Dice} and \ipie implement the same generalized Wick's theorem algorithm and therefore they both can handle many more determinants in the trial wavefunction.
We test our implementation on \ce{H50} and \ce{[Cu2O2]^{2+}} using the heat-bath configuration interaction (HCI) wavefunctions\cite{Holmes2016Aug,Smith2017Nov} made available in ref. \citenum{Mahajan2021Aug} as our trial. Note that a converter is provided with ipie to allow HCI wavefunctions generated with \texttt{Dice} to be used. We found our code to be slightly faster than \texttt{Dice}. The implementation in \texttt{QMCPACK}
is already 10 times slower for a modest determinant count ($10^3$). We expect that our implementation will be well suited for studying systems with modest strong correlation and large-scale dynamic correlation.

For GPU, it is often
the case that
one cannot store the auxiliary vectors on one GPU
due to the limited memory available on GPU.
In that case, we split auxiliary vectors into multiple blocks and distribute each block over multiple GPUs.
To minimize the size of communications during the run,
we communicate walker wavefunctions instead of auxiliary vectors as well as the result of computation such as local energy and force bias.
This multi-GPU implementation will become important when studying large systems. We hope to report applications that utilize this feature in the future.

\subsection{Isomerization energy of \ce{[Cu2O2]^{2+}}}
\ce{[Cu2O2]^{2+}} is one of the widely studied model problems and has been commonly referred to as the torture track.\cite{Cramer2006Feb}
Both strong and weak correlation play an important role in its electronic structure, and hence this system poses a great challenge to many of the existing electronic structure methods.
The computation of isomerization energies
between
bis($\mu$-oxo) and $\mu$-$\eta^2$:$\eta^2$ peroxo configurations
is often of interest in this problem.
These are motifs that can be found in enzymes such as tyrosinase.\cite{Solomon1992Jun}
These two structures are often characterized by
$f = 0$ and $f=1$, respectively, where $f$ is a structural parameter that allows one to interpolate between the two conformations by choosing $f\in[0,1]$.
In this work, we focus on computing the isomerization energy between these two structures.
The geometries used here have been taken from Ref. \citenum{Mahajan2021Aug}.
We use 640 walkers and $\Delta t$ of 0.005 a.u. for all calculations.  

\subsubsection{(32e,108o) correlation space}
Due to the difficulty of this problem,
it is necessary to employ
ph-AFQMC with an MSD trial based on HCI with self-consistent field (HCISCF) calculations. 
A recent unbiased QMC (i.e., fp-AFQMC)\cite{Mahajan2021Aug} study reported
statistically exact energies for this problem in double and triple-zeta bases.
For the smallest problem studied by fp-AFQMC (the double-zeta ANO basis denoted BS1 in Ref.~\citenum{Mahajan2021Aug}), we first assess
the accuracy of ph-AFQMC as a function of the number of determinants in our trial.
Following Ref.\citenum{Mahajan2021Aug} we freeze 10 core orbitals so that the trial wavefunction correlates 32 electrons and 108 orbitals.
\begin{figure}[!h]
    \centering
    \includegraphics{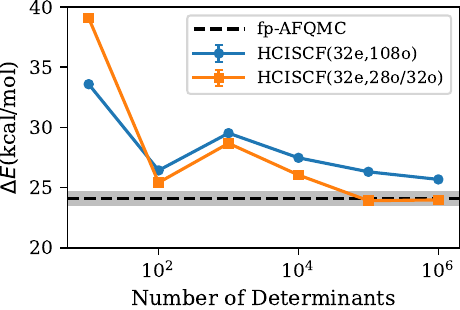}
    \caption{Convergence in $\Delta E = E(f=0)-E(f=1)$ for \ce{[Cu2O2]^{2+}} as a function of the number of determinants in HCISCF trial wavefunctions with two different types of active spaces (see main text for more details).
    The fp-AFQMC result was taken from Ref. \citenum{Mahajan2021Aug} and its error bar is indicated by the shaded region.
    The correlation space involves (32e, 108o) and the basis set used is BS1 given in Ref. \citenum{Mahajan2021Aug}.
    }
\label{fig:cu2o2bs1}
\end{figure}

In \cref{fig:cu2o2bs1}, we have two sets of ph-AFQMC results. They are using different HCISCF wavefunctions as trial wavefunctions.
One is using the HCISCF wavefunction generated for 32 electrons and 108 spatial orbitals, but the other one is using a trial generated for 32 electrons and 28 spatial orbitals ($f=0$) or 32 spatial orbitals ($f=1$).
For (32e,108o) HCISCF, we followed the procedure from Ref.~\citenum{Mahajan2021Aug} and used $\epsilon_1 = 10^{-3}$ (a relatively loose threshold\cite{Holmes2016Aug,Smith2017Nov} for orbital optimization, followed by a single-shot HCI calculation with $\epsilon_1 = 10^{-4}$.
This HCISCF trial with a larger active space shows a quite long tail in its convergence and even with $10^6$ determinants the isomerization energy is off by a kcal/mol or more. 
This points towards the poor quality of this HCI trial wavefunction.

To remedy this situation, we took a different strategy where we generate a reduced active space based on natural occupation numbers obtained from the HCI trial with a larger active space (in this case (32e,108o)). For the purpose of obtaining a smaller active space, we found that the HCI wavefunction for (32e,108o) active space with $\epsilon_1=10^{-4}$ is accurate enough. We included orbitals with occupation $n$ that satisfies $0.01\le n\le1.99$ in the subsequent, more precise HCISCF calculations.
These active spaces are 28-orbital and 32-orbital, respectively, for $f=0$ and $f=1$.
Using these new active spaces, we perform more accurate HCISCF calculations with $\epsilon_1 = 5\times10^{-5}$.
For these smaller active spaces, the variational HCISCF calculation is nearly converged in that the perturbative contribution within the active space is only 3-4 m$E_h$.
These new more compact HCISCF trials show quick convergence in the isomerization energy where our ph-AFQMC relative energies are within the error bar of the exact fp-AFQMC energies with $N_\text{det} = 10^5$ or more. 

\subsubsection{Larger correlation space}
Given the success of natural orbital active spaces in the previous section,
it is useful to extend this approach to
larger correlation space.
For a larger number of electrons and orbitals, even qualitatively correct HCISCF calculations become computationally infeasible.
Therefore, we chose to project the converged HCISCF orbitals in a smaller basis on to a larger basis. 
For BS1 (52e,108o), such a projection is not necessary as one can use the same trial wavefunction as the one used for (32e,108o).

For larger ANO-RCC bases\cite{Fdez.Galvan2019Nov} (ANO-RCC-VTZP; called BS2 in Ref. \citenum{Mahajan2021Aug}), 
we performed
HCISCF calculations ($\varepsilon_1=2.5\times10^{-5}$) with ANO-RCC-VDZP to obtain a compact active space based on natural orbital occupation numbers for both configurations.
For the larger basis sets we tightened the natural orbital threshold to $0.02 \le n \le 1.98$.
These active spaces are (32e,31o) and (32e,32o), respectively, for $f=0$ and $f=1$, similar to those in BS1.
The projection onto a larger basis set can be performed via
\begin{equation}
\mathbf C_\text{large}
=
\mathbf S_\text{large}^{-1} \bar{\mathbf S}_\text{large,small} \mathbf C_\text{small}
\end{equation}
where
$\mathbf C_\text{small}$
is the molecular orbital (MO) coefficient in the smaller basis set, 
$\mathbf C_\text{large}$
is the MO coefficient in the larger basis set,
$\mathbf S_\text{large}^{-1}$
is the inverse of the overlap matrix between atomic orbitals (AOs) in the larger basis set, and
$\bar{\mathbf S}_\text{large,small}$
is the overlap between AOs in large and small basis sets.\cite{Montgomery2000Apr}
After symmetrically orthogonalizing these projected MOs, we perform HCISCF calculations to relax orbitals further.

\begin{figure}[!h]
    \centering
    \includegraphics{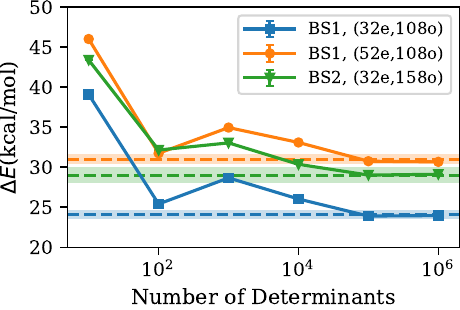}
    \caption{Convergence in $\Delta E$ ($E(f=0) - E(f=1)$) for \ce{[Cu2O2]^{2+}} as a function of the number of determinants in HCISCF trial wavefunctions for different correlation spaces and bases (BS1, BS2). The fp-AFQMC results taken from Ref. \citenum{Mahajan2021Aug} are shown as dotted lines and their error bars are indicated by shaded regions.
    }
\label{fig:cu2o2_larger}
\end{figure}

Using the strategy described above,
we performed 
larger correlation space calculations as shown in \cref{fig:cu2o2_larger}.
In both BS1 (52e,108o) and BS2 (32e,158o), 
we reach exact isomerization energies at a determinant count of $10^5$.
We note that 
ph-AFQMC absolute energies for both $f=0$ and $f=1$
are
converged better than 1 m$E_h$ with $N_\text{det}= 10^6$ from available fp-AFQMC total energies.
However, for $N_\text{det}= 10^5$, we are benefiting from cancellation of error. 

\subsubsection{Towards the basis set limit}
\begin{table}[!h]
\begin{tabular}{|c|r|r|r|}
\hline
\multicolumn{1}{|c|}{$N_\text{det}$} & \multicolumn{1}{c|}{TZ} & \multicolumn{1}{c|}{QZ} & \multicolumn{1}{c|}{CBS} \\ \hline
$10^4$                      & 30.4(2)                 & 27.0(2)                 & 28.1(3)                  \\ \hline
$10^5$                     & 29.0(4)                 & 25.4(3)                 & 26.3(5)                  \\ \hline
$10^6$                    & 29.1(6)                 & 25.2(4)                 & 26.0(7)                  \\ \hline\hline
fp-AFQMC\cite{Mahajan2021Aug} & 29(1) & N/A & N/A \\ \hline
\end{tabular}
\caption{
ph-AFQMC \ce{[Cu2O2]^{2+}} isomerization energies (kcal/mol) for ANO-RCC-VTZP (BS2), ANO-RCC-VQZP, and the complete basis set (CBS) limit for different determinant counts ($N_\text{det}$).
32 electrons were correlated in these calculations and the X2C Hamiltonian\cite{Liu2010Jul} was used for relativistic effects.
\label{tab:cu2o2}
N/A means ``not available.''
}
\end{table}

\begin{table}[!h]
\begin{tabular}{|c|r|r|r|}
\hline
\multicolumn{1}{|c|}{$N_\text{det}$} & \multicolumn{1}{c|}{TZ} & \multicolumn{1}{c|}{QZ} & \multicolumn{1}{c|}{CBS} \\ \hline
$10^4$ & 31.6(4) & 29.6(3) & 31.8(5) \\ \hline
$10^5$ & 29.1(5) & 27.2(4) & 29.5(6)\\ \hline
$10^6$ & 29.0(7) & 27.1(9) & 29(1)\\ \hline
\end{tabular}
\caption{
Same as \cref{tab:cu2o2} except that 52 electrons are correlated.
\label{tab:cu2o252e}
}
\end{table}

\begin{table}[!h]
\begin{tabular}{c|c|r|r}
\hline
\multicolumn{1}{c|}{Method} &
\multicolumn{1}{|c|}{Basis} & \multicolumn{1}{c|}{Correlation Space} & \multicolumn{1}{c}{$\Delta E$} \\ \hline
fp-AFQMC & TZ & (32e, 158o) & 29(1) \\ 
ph-AFQMC & & (32e, 158o) & 29.1(6) \\ 
CCSD(T)  & & (32e, 158o) &  34.8\\ 
\hline
ph-AFQMC & TZ & (52e, 166o) & 29.0(7) \\ 
CCSD(T)  & & (52e, 166o) &  33.6\\ 
\hline
ph-AFQMC & QZ & (52e, 290o) & 27.1(9) \\ 
CCSD(T)  & & (52e, 290o) &  33.0\\ 
\end{tabular}
\caption{
Comparison between ph-AFQMC($N_\mathrm{det}=10^6)
$ fp-AFQMC and CCSD(T) for different basis set and correlation space sizes. CCSD(T) and fp-AFQMC results are from Ref. ~\citenum{Mahajan2021Aug}. Note that the fp-AFMQC result for BS3 are actually the (32e, 158o) results from BS2 with a basis set correction from CCSD(T) added\citep{Mahajan2021Aug}, whereas the fp-AFQMC results were run in the full correlation space.
For comparison purposes, the HSCISCF trial wavefunction isomerization energies are 59.2 and 55.7 kcal/mol for BS2 and BS3 respectively.
\label{tab:cu2o2_comparison}
}
\end{table}

We further apply our ph-AFQMC implementation to computing the isomerization energy towards the complete basis set (CBS) limit.
ANO-RCC-VQZP involves correlation space of (32e,280o) and (52e,290o).
We computed the correlation energy of TZ and QZ using the mean-field energy computed by the dominiant determinant. These correlation energies were extrapolated to the CBS limit using $1/Z^3$ where $Z$ is the cardinality of basis sets.\citep{helgaker1997basis,halkier1998basis}
We then combine the CBS limit correlation energy with the QZ mean-field energy to compute the final CBS total energy.

As shown in \cref{tab:cu2o2}, for ANO-RCC-VQZP, the ph-AFQMC isomerization energy changes less than 1 kcal/mol going from $10^5$ to $10^6$ determinants similarly to ANO-RCC-VTZP.
Similarly, our CBS estimates also show a small change between $10^5$ and $10^6$ determinants.
In the CBS limit, our best estimate for the isomerization energy with 32 electrons correlated is 26.0(7) kcal/mol.
Given the comparison against fp-AFQMC for small bases and the energy change smaller than 1 m$E_h$ between $10^5$ and $10^6$ determinants, we think our ph-AFQMC CBS correlation energy for both $f=0$ and $f=1$ is converged better than 1 m$E_h$.
In \cref{tab:cu2o252e}, we present a similar result where 52 electrons are correlated. While we observe quantitative differences between \cref{tab:cu2o2} and \cref{tab:cu2o252e}, the qualitative behavior as a function of $N_\text{det}$ is similar. Our best theoretical estimate for the isomerization with 52-electron correlation space is 29(1) kcal/mol.
We expect our estimate of the isomerization energy for 32- and 52-electron correlation space in the CBS limit to be 
nearly within its statistical error bar from the exact theoretical answer.
We emphasize that such reliable high-level calculations have never been possible with other single- and multi-reference methods such as density matrix renormalization group \cite{Yanai2010Jan,phung2016cumulant} or high-order coupled-cluster theory in such large correlation spaces.\cite{Cramer2006Feb,Malmqvist2008May,zou2022efficient}
Finally, \cref{tab:cu2o2_comparison} provides some comparison between ph-AFQMC, ph-AFQMC and CCSD(T) in these larger correlation spaces.

\section{Conclusions}
In this paper, we report
a Python implementation of ph-AFQMC, \texttt{ipie},
that can efficiently run on both CPU and GPU.
\texttt{ipie} offers flexible development environment utilizing Python infrastructure as well as sufficient performance for production use. 
In \texttt{ipie}, we make an extensive use of NumPy for CPU and CuPy for GPU
and sometimes rely on Numba's \texttt{jit} capability for certain computational kernels.
Our preliminary implementation achieves similar or better CPU and GPU performance compared to other C++ codes such as \texttt{QMPACK} and \texttt{Dice}.
For CPU, our code significantly accelerates prototyping new theoretical developments pertinent to ph-AFQMC due to the flexibility provided by Python.
For GPU, we only had to write a few customized CUDA kernels. For instance, for SD trials, we only needed one customized CUDA kernel written with Numba's \texttt{cuda.jit} for the local energy evaluation.
This is in contrast to Refs. \citenum{Shee2018Aug,Malone2020Jul}, which were 
written in C++. These C++ implementations relied on many customized CUDA kernels which we could avoid largely due to CuPy handling complex tensor operations.

Using \texttt{ipie}, we also reported new ph-AFQMC results using HCISCF trials on \ce{[Cu2O2]^{2+}}
where ph-AFQMC isomerization energies could be converged to exact answers whenever
these are available from unbiased fp-AFQMC.\cite{Mahajan2021Aug}
Furthermore, we report isomerization energies with QZ and in the basis set limit.
The largest correlation calculation involves $10^6$ determinants in trial, 52 electrons, and 290 orbitals.
We expect our isomerization energies are better than 1 kcal/mol from the exact answer given the relatively small changes
observed when increasing the number of determinants in our trial.
Since \ce{[Cu2O2]^{2+}} is an example where ph-AFQMC with other trial wavefunctions is significantly inaccurate\cite{LandinezBorda2019Feb} and other
multi-reference perturbation theory methods fail,
our results
highlight the utility of ph-AFQMC
with HCISCF trials.

We plan to expand our code's capability to finite-temperature ph-AFQMC,\cite{Lee2021Feb} electron-phonon problems,\cite{Lee2021Mar} lattice Hamiltonians, and other QMC methods in the future.
We hope that our open-source package, \texttt{ipie} (\href{https://github.com/linusjoonho/ipie}{https://github.com/linusjoonho/ipie}), will gain larger user and developer bases
and will ultimately serve as a community code for prototyping and applying ph-AFQMC in quantum chemistry. 

\section{Acknowledgements}
We thank Hung Pham and Nick Rubin for initial involvement.
J.L. thanks David Reichman and Ryan Babbush for encouragement and support.
We thank Soojin Lee for the \texttt{ipie} logo design and illustration.
The work of J.L. was supported in part by a Google research award.

\section{Data Availability}
The data that support the findings of this study are openly
available in our Zenodo repository at \href{https://doi.org/10.5281/zenodo.7061987}{https://doi.org/10.5281/zenodo.7061987}.

\bibliography{refs}

\end{document}